\documentclass[aps,pre,reprint,amsmath,amssymb,groupedaddress,showkeys]{revtex4-1}
\pdfoutput=1

\usepackage{graphicx}
\usepackage{amsmath,amssymb}
\usepackage{siunitx, microtype, xspace, hyperref, parskip, xcolor}
\usepackage{booktabs, threeparttable}

\usepackage{tikz}
\usetikzlibrary{positioning, arrows}
\tikzstyle{every picture}+=[remember picture]

\usepackage{pgfplots}
\pgfplotsset{compat=newest}
\usepgfplotslibrary{units}
\usepgfplotslibrary{external}
\usetikzlibrary{pgfplots.external}
\usetikzlibrary{pgfplots.colormaps}
\graphicspath{ {./img/} }

\begin{document}

\title{Balanced Information Storage and Transfer in Modular Spiking Neural Networks}

\author{Pedro A.M. Mediano}
\email{pmediano@imperial.ac.uk}
\author{Murray Shanahan}
\affiliation{Department of Computing, Imperial College London}
\date{\today}

\begin{abstract}

  While information processing in complex systems can be described in abstract,
  general terms, there are cases in which the relation between these
  computations and the physical substrate of the underlying system is itself of
  interest. Prominently, the brain is one such case. With the aim of relating
  information and dynamics in biological neural systems, we study a model
  network of spiking neurons with different coupling configurations, and
  explore the relation between its informational, dynamical, and topological
  properties. We find that information transfer and storage peak at two
  separate points for different values of the coupling parameter, and are
  balanced at an intermediate point. In this configuration, avalanches in the
  network follow a long-tailed, power law-like distribution. Furthermore, the
  avalanche statistics at this point reproduce empirical findings in the
  biological brain.

\end{abstract}


\maketitle

\section{Introduction}
\label{sec:intro}

Information theory has been an invaluable tool for neuroscience, and in the
past few decades it has been making great contributions to our understanding of
neural computation and coding \cite{Dayan2001}. This has inaugurated a whole
research field bringing together both disciplines \cite{Borst1999,Johnson2001}.
The broad goal of this research is to describe neural computation in abstract
terms, to dissociate the cognitive process from its neural implementation. This
has proven to be a fruitful and interesting endeavour, since an abstract
account would allow us to compare the brain with other cognitive systems, both
biological and artificial.

However, in deliberately ignoring the physical substrate of computations, a
purely information-theoretic view of the brain misses some interesting research
questions: what kinds of dynamical states lead to what kinds of computations?
Does a particular process make use of all resources available to the neurons?
Could a given computation be instantiated by a different dynamical process? To
address these and other questions we must consider the specific mechanisms by
which the neural physical substrate gives rise to emergent computations.

For these reasons we advocate a hybrid view of neural computation, in which
information and dynamics are two sides of the same coin \cite{Beer2014}. Along
these lines, several authors have established connections between
information-theoretic and dynamical properties of neural networks at several
scales: specific single-neuron-level mechanisms have been found to be
informationally optimal in some sense\cite{Lochmann2008,Hennequin2010}, and on
a larger scale criticality has been linked to increased information transfer
\cite{DeArcangelis2010} and information capacity \cite{Shew2011}.

In this article we continue this line of research by linking information,
dynamics and topology in a modular network of spiking neurons
\cite{Shanahan2008}. Specifically, we investigate the relation between
dynamical criticality, aspects of information storage and transfer, and the
balance between local and global coupling, matching the findings in our model
with observed experimental data.

We find that information transfer and storage peak at two separate points for
different values of the coupling parameter, and are balanced at an intermediate
point. In this configuration, avalanches in the network follow a long-tailed,
power law-like distribution. Furthermore, the avalanche statistics at this
point reproduce empirical findings in the biological brain \cite{Beggs2003}.

\section{Methods}
\label{sec:methods}

We consider a system similar to the one shown in Ref.~\cite{Shanahan2008}. The
network consists of a total of 1000 neurons, comprising one population of 200
inhibitory neurons and $n=8$ populations (or \emph{modules}) of 100 excitatory
neurons each.

A total of 1000 internal one-directional connections (or \emph{synapses}) are
added to each excitatory module, such that any given pair of neurons are
connected with probability $0.1$ -- resulting in modules of 10\% edge density.
Synapses from excitatory to inhibitory neurons are focal, with every 4
excitatory neurons in the same module projecting to the same inhibitory neuron.
Every inhibitory neuron is connected to all other neurons in the network. The
delay of each excitatory-excitatory synapse is sampled at random from the
$[1,20]$\si{\milli \second} interval, and the delay of all other synapses is
fixed at \SI{1}{\milli \second}.

Once initialised, the network is subject to a \emph{rewiring process}, akin to
the one proposed by Watts and Strogatz \cite{Watts1998}. Watts and Strogatz's
key result is that the network undergoes a transition regime in which strong
clustering coexists with short path lengths, making the network simultaneously
segregated and integrated -- termed a `small-world' network. Here we seek to
investigate how such small-world topological properties affect the dynamical
and informational behaviour of the network.

\begin{figure}[ht]
  \centering
  \includegraphics{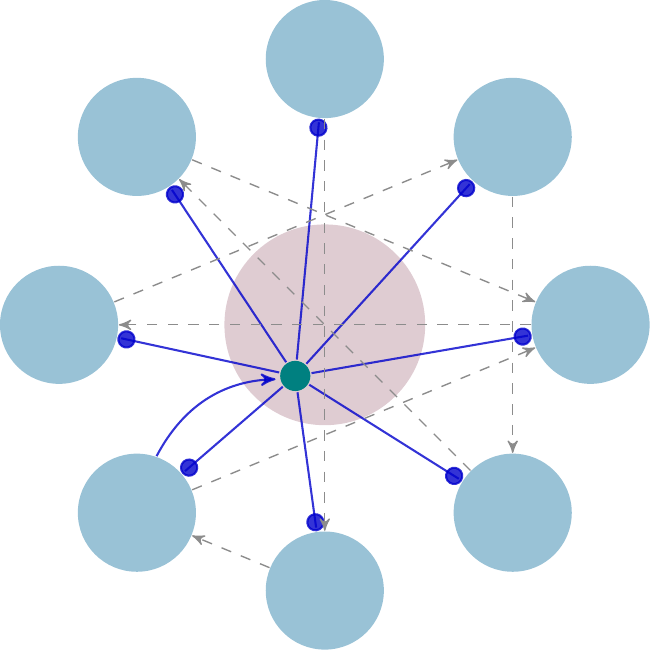}
  \caption{Schematic diagram of the model network. There are 8 excitatory
    modules (light blue) connected to one another and to a larger inhibitory
    pool (light purple). Inhibitory neurons have diffuse connections to all the
    network (blue round arrows), and excitatory neurons have focal connections
    to inhibitory neurons (blue pointed arrow) and long-range connections
  between them (dashed gray arrows).}
  \label{fig:netDiagram}
\end{figure}

The rewiring process is only applied to the 800 excitatory neurons, and is
implemented as follows. With probability $p$, each synapse is detached from its
target neuron, and assigned a new target picked uniformly at random from any
excitatory module, thus introducing inter-module synapses. This rewiring
probability $p$ effectively regulates the balance between local intra-module
coupling and long-range inter-module coupling, and is the main object of
analysis in this paper.

Once the topology of the network is set, we add a dynamical model to simulate
the spiking behaviour of biological neurons. The dynamics of each neuron are
simulated using the Izhikevich model \cite{Izhikevich2003},
\begin{subequations}
  \begin{align}
    \frac{dv}{dt} & = 0.04 v^2 + 5 v + 140 - u + I \\
    \frac{du}{dt} & = a (b v - u) ~ ,
  \end{align}
  \label{eq:iz1}%
\end{subequations}
\noindent where $v$ is the membrane potential (or voltage) of the neuron, $u$
is an auxiliary recovery variable and $I$ is the incoming current from ingoing
synapses or external sources. All quantities are in arbitrary units. When the
voltage of any given neuron goes above a certain threshold we record a discrete
\emph{spike} event, such that
\begin{align}
  \mathrm{if~} v \geq 30 \mathrm{, then}
  \begin{cases} v \leftarrow c \\ u \leftarrow u + d ~ . \end{cases}
  \label{eq:iz2}%
\end{align}
The values for the $a,b,c,d$ parameters for both excitatory and inhibitory
neurons are taken verbatim from Ref.~\cite{Izhikevich2003}. We note that
neurons in all populations are slightly heterogeneous, as neuron parameters are
randomised.

Once the network topology and the neuron parameters are set, the network can be
simulated by numerically integrating Eqs.~\eqref{eq:iz1} and \eqref{eq:iz2}. We
store all the spiking events from excitatory neurons for future analysis and
ignore the spikes in the inhibitory population.

\subsection{Information Theory}
\label{sec:it}

Information Theory (IT) is a useful and increasingly popular tool to analyse
complex systems \citep{Prokopenko2009,Lizier2011,Mediano2016}. We refer the
reader to the classic textbook by Cover and Thomas for a general introduction
to IT \cite{Cover2006}.

Throughout the paper we denote the random variable representing the activity of
all excitatory neurons at time $t$ by $S_t$. The system is partitioned in 8
parts, corresponding to the 8 modules of excitatory neurons described above.
The time series of module $i$ is denoted by $M_{i, t}$. We reserve the symbols
$X,Y,Z$ for arbitrary random variables.

A cornerstone of IT, Mutual Information (MI) can be used to quantify
interdependence between two random time series $X_t, Y_t$. More precisely, it
measures the reduction in uncertainty about $X_t$ that results from observing
the value of $Y_t$ (or viceversa). One of the many ways to define mutual
information is
\begin{align}
  \mathrm{MI}(X_t, Y_t) =  H(X_t) + H(Y_t)  - H(X_t, Y_t) ~ ,
  \label{eq:mi}%
\end{align}
\noindent where $H$ is the standard Shannon entropy. Note that MI is a
bivariate measure. To quantify how much information is shared between all $n=8$
parts of our system we use multiinformation (also called \emph{total
information}, TI), one possible multivariate extension of MI
\cite{Studeny1999}. TI is defined based on Eq.~\eqref{eq:mi} as
\begin{align}
  \mathrm{TI}(S_t) = \sum_{i=1}^n H(M_{i,t}) - H(S_t) ~ .
  \label{eq:ti}%
\end{align}
Our study adopts the framework for information processing in complex systems
introduced by Lizier \cite{Lizier2010}. According to Lizier and others,
distributed information processing in complex systems is the interaction
between three processes: information storage, information transfer and
information modification. A sound, rigorous account of information modification
is still an open problem, so we will restrict our analysis to storage and
transfer.

We first study information storage, following Ref.~\cite{Lizier2012}. We define
the entropy rate of a time series $H_\mu(X_t)$ as the entropy generated by one
single timestep of the series, given all its previous history -- i.e.
\begin{align}
  H_\mu(X_t) = \lim_{k \rightarrow \infty} H(X_t | X_{t-1}, X_{t-2}, \cdots X_{t-k}) ~ .
  \label{eq:entRate}%
\end{align}
This quantity measures the amount of information in the time series that is
fundamentally unpredictable; and enables us to write the following nonnegative
decomposition of $H(X_t)$:
\begin{align}
  H(X_t) = \mathrm{AIS}(X_t) + H_\mu(X_t) ~ ,
  \label{eq:entDecomposition}%
\end{align}
\noindent where the quantity of interest is AIS, the \emph{active information
storage}. This decomposition is intuitively interpretable: the information
needed to predict step $t$ in the time series is the information stored in its
entire previous history, plus the new information being generated. In other
words, AIS quantifies how much information about the history of the system is
useful in predicting the system's next state. Importantly for our purposes, it
has also been proposed as a tool to understand distributed computation in
neural and complex systems \cite{Wibral2014a}. As the $k \rightarrow \infty$
limit is (for obvious reasons) intractable, we write the finite-$k$
approximation of the AIS of module $i$ as
\begin{align}
  \mathrm{AIS}_k(M_{i,t}) = \mathrm{MI}(M^{(k)}_{i,t}, M_{i,t+1}) ~ ,
  \label{eq:ais}%
\end{align}
\noindent where $X^{(k)}_t$ is the $k$-dimensional embedding vector of $X$ at
time $t$, that contains the past $k$ values of $X$ up to and including time
$t$. The aim of this embedding vector is to capture the state of the underlying
dynamical process, and can be viewed as a state-space reconstruction in the
Takens sense \cite{Takens1981}.

To further understand the informational properties of the network, we measure
information transfer with \emph{transfer entropy} (TE) \cite{Schreiber2000}. TE
quantifies to what extent knowledge of $X_t$ contributes to predicting the
future of $Y_t$ beyond the information provided by the past of $Y_t$ alone, and
it is defined as
\begin{align}
  \mathrm{TE}_k(X \rightarrow Y) = \mathrm{MI}(X^{(k)}_t, Y_{t+1} | Y^{(k)}_t) ~ .
  \label{eq:te}%
\end{align}
There is a great body of theory behind TE as a measure of information transfer,
and it is closely related to causality in the Wiener-Granger sense
\citep{Lizier2010b,Barnett2012}. However, when there are more variables in the
system apart from $X$ and $Y$, TE does not capture exactly the immediate
influence of $X$ on $Y$ -- there may be other higher-order interactions,
mediated by other variables in the system \cite{Lizier2008}. For example, there
could be another variable $Z$ influencing both $X$ and $Y$ at different
timescales, which could distort the measurement. To measure exclusively the
direct transfer from $X$ to $Y$ without the influence of $Z$, we define the
\emph{conditional transfer entropy} as
\begin{align}
  \mathrm{CTE}_k(X \rightarrow Y | Z) = \mathrm{MI}(X^{(k)}_t, Y_{t+1} | Y^{(k)}_t, Z^{(k)}_t) ~ .
  \label{eq:cte}%
\end{align}
Finally, we use CTE to define a nonparametric version of \emph{causal density}
(CD), slightly different from the conventional formulation \citep{Seth2011}. We
define CD as the average pairwise CTE conditioned on the rest of the system,
i.e.
\begin{align}
  \mathrm{CD}_k(S_t) = \frac{1}{n (n-1)} \sum_{ij} \mathrm{CTE}_k(M_{i,t} \rightarrow M_{j,t} | S^{[ij]}_t) ~ ,
  \label{eq:cd}
\end{align}
\noindent where $S^{[ij]}_t$ represents the whole system $S_t$ with variables
$M_{i,t}$ and $M_{j,t}$ removed. This formulation is equivalent to the
conventional one (up to a constant) if all variables are Gaussian-distributed
\cite{Barnett2009}. Within this framework, we interpret CD as a global average
measure of information transfer.

From a different perspective, CD can be also thought of as quantifying
dynamical complexity \citep{Tononi1998}, an early branch of research in complex
neural systems now embodied in Integrated Information Theory
\citep{Balduzzi2008a}. A system is said to have high dynamical complexity if it
displays a balance of segregation (to the extent that its parts behave
independently) and integration (to the extent that the whole system acts as
one). In a completely segregated system in which the elements act independently
there can be no information transfer and CD is trivially null. On the other
end, in a completely integrated system in which all elements are heavily
correlated, element $i$ will provide no additional information about $j$ beyond
the information provided by the rest of the system $X^{[ij]}$, and CD is again
null. Thus, CD also provides a principled measure of dynamical complexity.

\section{Results}
\label{sec:results}

We generate 400 networks with different values of $p$ sampled uniformly at
random in the $[0, 1)$ interval, and another 200 with $p$ sampled exponentially
at random in the $(10^{-2}, 1)$ interval. This is to have dense coverage of the
parameter space at the low end of the range. The activity of each network is
simulated using the NeMo library \citep{Fidjeland2009} for \SI{200}{\second}
using the RK4 method with a timestep of \SI{0.2}{\milli \second}, and
subsampled to a resolution of \SI{1}{\milli \second}. The first \SI{1}{\second}
of simulation is discarded to avoid transient effects. Information-theoretic
quantities are calculated using the implementation in \cite{Lizier2014} and are
reported in bits.

\begin{figure*}
  \centering
  \includegraphics{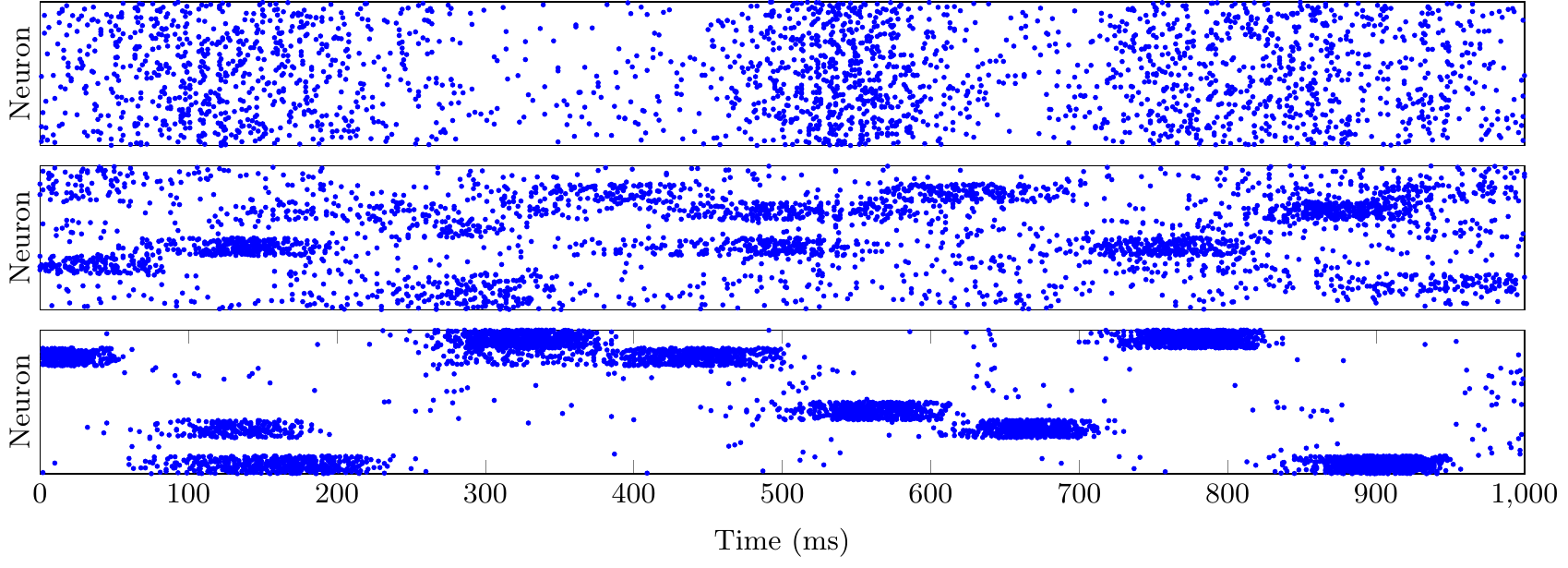}
  \caption{Sample runs of the network for different values of the rewiring
  probability $p$. The values are $p = 0.9$ (top), $0.2$ (middle) and $0$
  (bottom). As $p$ increases the system transitions from multistable
  competitive dynamics to oscillatory cooperative dynamics.}
  \label{fig:sampleData}
\end{figure*}

\subsection{Model Behaviour}

In this section we give a qualitative summary of the behaviour of the model
that will help interpretation of quantitative findings described in the rest of
the paper. Spike raster plots of representative runs of the network with
different values of $p$ are shown in Fig.~\ref{fig:sampleData}.

We begin with the fully modular network, the $p=0$ case (bottom panel in
Fig.~\ref{fig:sampleData}). In this setting there are no direct connections
between the excitatory modules. When any neurons in an excitatory module become
active, the high density of intra-module synapses ensures that all neurons in
the module quickly become activated.

Through the focal excitatory-inhibitory synapses, the active module feeds
charge to the subset of inhibitory neurons assigned to it. These start spiking
rapidly, and because of the diffuse connections they shut down the activity in
all other excitatory modules. This results in competitive multistable dynamics,
as one module gaining control of the network prevents all others from doing so.
In computational neuroscience this kind of competition mechanism is known as
Winner-Take-All (WTA). Subsequently, the active module saturates and the
refractory period of the neurons makes it cease firing, so that other module
can take over.

At the other end of the parameter range, at $p=0.9$ (top panel in
Fig.~\ref{fig:sampleData}), the dynamics are very different. Topologically,
this setting corresponds to a fully random, Erd\H{o}s-R\'enyi network. There is
no notion of modules anymore, and all excitatory neurons are statistically
equivalent. The result is an interaction between a uniform population of
excitatory neurons with a smaller group of inhibitory neurons. This is
reminiscent of a known mechanism of oscillation generation -- a PING
architecture \cite{Buzsaki2012}. The interplay between excitation and
inhibition and the synaptic delays between them make the whole system
oscillate. In this regime the modules are strongly correlated and cooperate in
maintaining the global oscillation.

Finally, at intermediate values of $p$ (middle panel in
Fig.~\ref{fig:sampleData}), these two opposite trends coexist. The dynamics of
the system are more chaotic and there is no clear pattern. Local and long-range
coupling are balanced and both affect the emergent dynamics (we recall that the
total number of connections is fixed, so an increase in long-range coupling is
always at the expense of a weaker intra-module coupling).

This transition is also interpretable as an emergent synchronisation
phenomenon. For low $p$ the WTA mechanism pushes the modules out of phase, and
the network is maximally desynchronised. Conversely, for high $p$ the modules
blend together and the synchrony between them increases.

In summary, the model we described features a transition from a competitive to
cooperative regime, controlled by a continuous parameter. This model naturally
interpolates between two neural circuits ubiquitously present in the cortex:
PING oscillators and multistable WTA circuits. As we describe below, it is
between these two extremes where critical dynamics and complex information
processing take place.

\subsection{Avalanche Statistics}

The seminal work of Beggs and Plenz \cite{Beggs2003} set out the search for
criticality in neural systems, in particular through the analysis of avalanche
statistics. A \emph{neural avalanche} is defined as a period of continued
spiking activity -- i.e. a period in which the activity of a neural population
is continuously above a certain \emph{avalanche threshold}. The \emph{avalanche
size} is the total number of spikes fired by all neurons in the population
between any two points of below-threshold activity.

By counting occurrences of avalanches in the network and recording their size
we obtain the \emph{avalanche size distribution}, a very relevant mathematical
construct subject of much study in statistical physics and complex systems
research. A common signature of critical dynamics and phase transitions is that
avalanche sizes follow a \emph{power law distribution} \cite{Pruessner2014},
defined as
\begin{align}
  P(s) \propto s^{-\alpha} ~ ,
  \label{eq:powerlaw}%
\end{align}
\noindent where $\alpha$ is called the \emph{critical exponent}. Beggs and
Plenz's key result is that measured activity in the biological brain
consistently follows a power law avalanche size distribution -- leading to the
hypothesis that the brain operates in a critical regime. Although their claims
on criticality have been contested \cite{Priesemann2014}, their empirical
finding of power law distributions in neural recordings is widely accepted.

In this section we present the avalanche analysis of the resulting activity of
our model. As a general methodological note, we mention that estimating and
evaluating power laws when working with empirical data is remarkably
complicated. In this analysis we use the methods and implementation provided in
\cite{Clauset2007, Alstott2013}.

\begin{figure}[ht]
  \centering
  \includegraphics{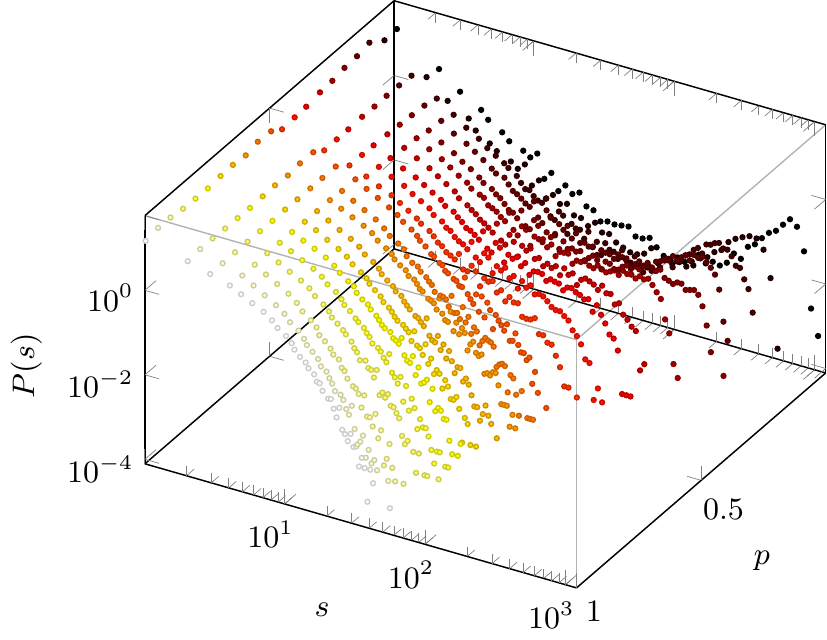}
  \caption{Avalanche size distributions for different rewiring probabilities
  $p$. As connections become delocalised, the system shifts from supercritical
  to subcritical. The $p$-axis is reversed for visualisation purposes.}
  \label{fig:avHist}
\end{figure}

We generate and run networks for many values of $p$ as described above and
measure avalanches in each module. To do this we calculate the mean firing rate
of each module over \SI{1}{\milli \second} bins and run the analysis with an
avalanche threshold of 3 spikes/ms. The distributions of the 8 modules in the
same run of the experiment are aggregated together to improve statistics.
Log-log avalanche size histograms are shown for evenly spaced values of $p$ in
Fig.~\ref{fig:avHist}.

For low values of $p$, when connections are highly localised, the system is
supercritical -- the avalanche size distribution is characterised by a
prominent peak at the far tail, which indicates that a disproportionately large
fraction of the avalanches are strongly energetic and saturate the modules.

Conversely, at high values of $p$ the system is subcritical. Avalanches are
weak and the avalanche size distribution has a short exponential tail. This is
probably caused by the diffuseness of the connectivity pattern -- rewiring
keeps the global synaptic strength fixed, but the influence of each burst of
activity is spread across the whole network instead of focalised in one single
module.

\begin{figure}[ht]
  \centering
  \includegraphics{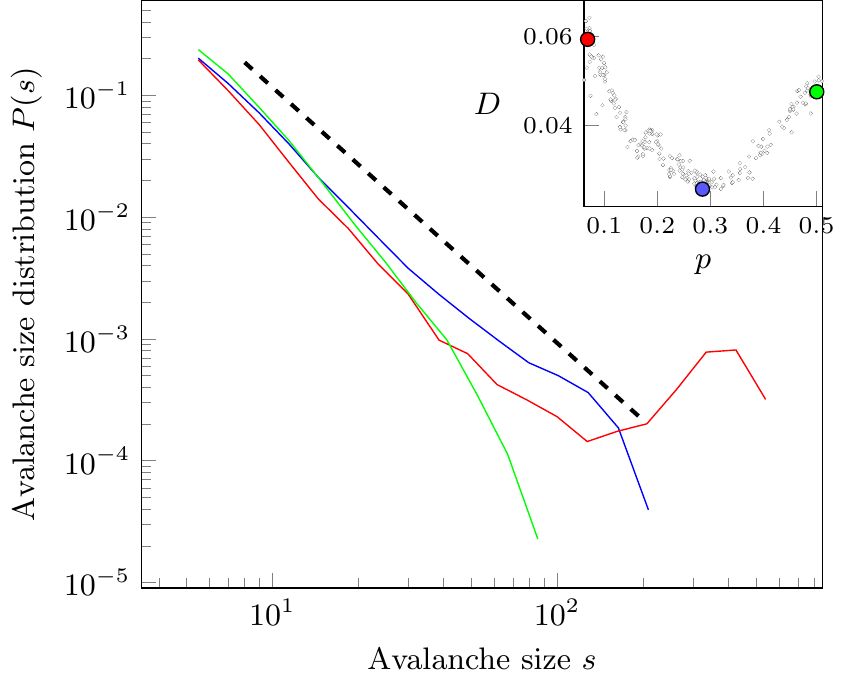}

  \caption{ Avalanche size distributions for three runs of the simulation,
    supercritical (\textcolor{red}{red}),  subcritical
    (\textcolor{green}{green}) and critical (\textcolor{blue}{blue}). Black
    line: reference $\alpha=-2$ power law as reported by
    \cite[Fig.~3A]{Beggs2003} for \SI{1}{\milli \second}-binned LFP data.
    Inset: Kolmogorov-Smirnov statistic $D$ comparing the data against a
    theoretical power law with the estimated parameters.  Filled circles in the
    inset correspond to the runs shown in the main plot.  }

  \label{fig:ksInset}
\end{figure}

It is at middle that the activity of the modules resembles the activity of a
critical system. Avalanche size distributions show power law-like statistics,
with a characteristic straight line in the log-log histogram and a small
protuberance at the end, result of finite-size effects. To test the claim that
the behaviour of the system is closest to a power law at an intermediate value
of $p$, we perform a maximum-likelihood power law fit to each trial and
calculate the 1-sample Kolmogorov-Smirnov (KS) statistic between the measured
data and the fitted power law. The results, together with three representative
histograms are shown in Fig.~\ref{fig:ksInset}.

This figure more clearly shows the difference between critical, subcritical and
supercritical behaviour; and the KS statistic determines that at $p=0.3$ the
system's avalanche size distribution is closest to a power law. Furthermore, at
that point the critical exponent of the maximum-likelihood fit is consistent
with the $\alpha \approx -2$ value found by Beggs and Plenz for \SI{1}{\milli
\second}-binned LFP data and by de Arcangelis and Herrmann in simulations of
realistic scale-free network topologies \cite{Beggs2003,DeArcangelis2010}.

\subsection{Information-theoretic analysis}

In practice, the challenge behind computing information-theoretic measures
amounts to estimating probability densities for the involved quantities (e.g.
$p(S_t)$ and $p(M_{i,t})$ in the case of TI). For our analyses we use the
nearest-neighbour estimators devised by Kraskov, St\"ogbauer and Grassberger
\cite{Kraskov2004}. The KSG estimators are non-parametric and make only weak
assumptions on the local neighbourhoods of the estimated probability density,
which makes them a robust, flexible tool. Reported results are corrected with
surrogate data methods \cite{Lucio2012}.

\begin{figure}[ht]
  \centering
  \includegraphics{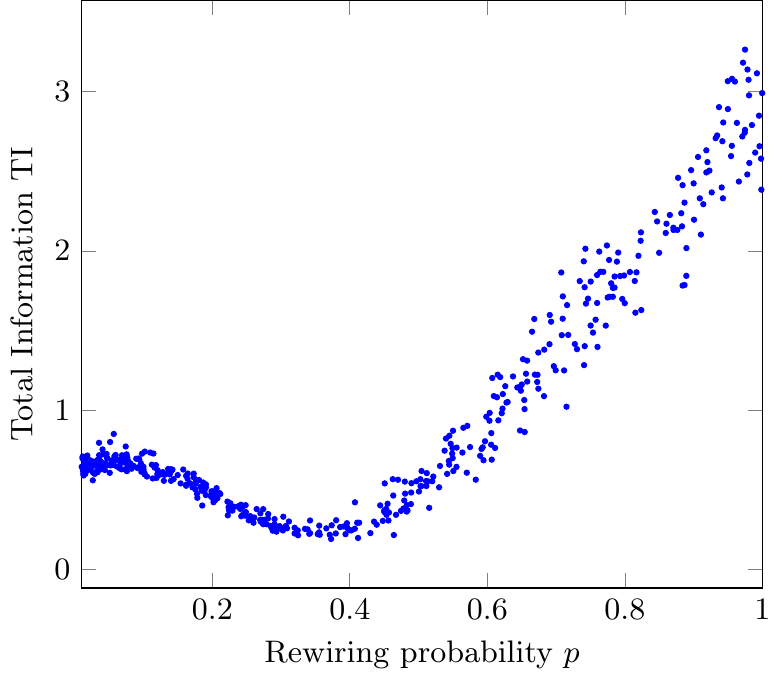}

  \caption{Total Information (TI) between all 8 excitatory modules in the
  network. TI is maximal in the high-$p$ regime, when modules are
indistinguishable. In the middle, where the topology of the network is more
complex, TI drops as the dynamics become more chaotic.}

  \label{fig:multiinfo}
\end{figure}

First, we show in Fig.~\ref{fig:multiinfo} the total information TI between all
8 modules of the network. As $p$ becomes large, the modules blend together and
the system's TI increases, reflecting the increase in instantaneous correlation
and synchronisation between the modules. This is evidence for cooperation
between the modules in the high-$p$ regime. Conversely, for $p=0$ the modules
are disconnected and TI is much lower. Note, however, that the modules have
information about each other because they still interact indirectly through the
inhibitory pool.

\begin{figure}[ht]
  \centering
  \includegraphics{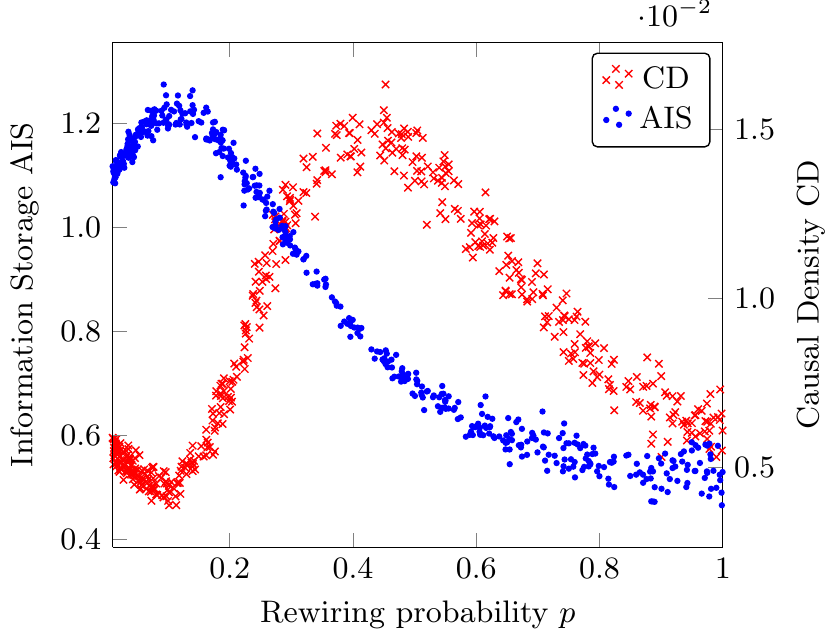}

  \caption{Active information storage (\textcolor{blue}{blue}, left axis) and
    nonparametric causal density (\textcolor{red}{red}, right axis) for
    different rewiring probabilities $p$. Each measure peaks at one side of the
    critical region around $p=0.3$ where the system shows power law-like
    statistics (see Fig.~\ref{fig:ksInset}).}

  \label{fig:tedensity}
\end{figure}

More importantly, we measure information storage and transfer with AIS and CD
and show the results in Fig.~\ref{fig:tedensity}. AIS is calculated separately
for each module and then the 8 modules are averaged for each run. Nonparametric
CD is calculated as described in section \ref{sec:it}. The embedding dimension
$k$ is fixed at 5 for all calculations.

First we note that information storage dominates the low-$p$ regime. Because of
the WTA competition mechanism, if a module is inactive it tends to remain
inactive, whereas if it is active it will most likely saturate and cease
activity shortly after. This means that the recent history of the module's
activity is highly informative of their future.

Regarding transfer, CD has a prominent peak in the mid-$p$ region. As expected,
there is little transfer in the $p=0$ or $p=1$ extremes, away of the
neighbourhood around the critical transition. This is because in the low-$p$
regime the modules are completely disconnected; and in the high-$p$ regime the
modules are so correlated that module $i$ no longer provides information about
$j$ after conditioning on $S^{[ij]}$.

More interesting is the neighbourhood around $p=0.3$, where storage and
transfer are maximally balanced. This coincides with the point where the
avalanche dynamics are closest to a power law, as measured by the KS statistic
and shown in Fig.~\ref{fig:ksInset}. This suggests that there is a
configuration of the system in which the balance between local and global
coupling results in a balance between local information storage and long-range
information transfer, which moreover is accompanied by a near-critical
avalanche distribution. This finding links together three complementary views
on neural computation: topological, informational and dynamical complexity.

\subsection{Criticality and linear interactions}
\label{sec:linear}

\begin{figure*}
\centering

\pgfplotsset{
  scale only axis,
  scaled ticks=false,
  enlarge x limits=0,
  enlarge y limits=0,
  width=0.9\linewidth,
  xtick=\empty,
  ytick=\empty,
  colormap={hot2}{[1cm]rgb255(0cm)=(0,0,0) rgb255(3cm)=(255,0,0) rgb255(6cm)=(255,255,0) rgb255(8cm)=(255,255,255)}
  }
\pgfkeys{/pgf/number format/fixed}
\centering

\newlength\figureheight
\newlength\figurewidth
\newlength\figurevsep
   \setlength\figureheight{1.8cm}
   \setlength\figurewidth{0.18\linewidth}
   \setlength\figurevsep{0.5em}

\begin{tabular}{c c c c c c}

  ~ & $p = 0.0$ & $p = 0.2$ & $p = 0.4$ & $p = 0.6$ & $p = 0.8$ \\

  D0 &
  \raisebox{-.5\height}{\includegraphics{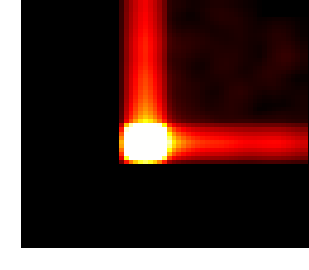}} &
  \raisebox{-.5\height}{\includegraphics{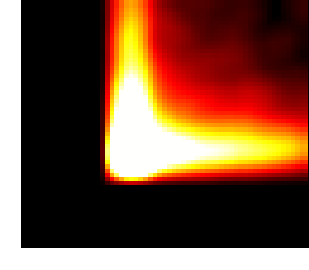}} &
  \raisebox{-.5\height}{\includegraphics{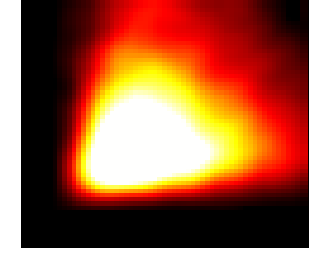}} &
  \raisebox{-.5\height}{\includegraphics{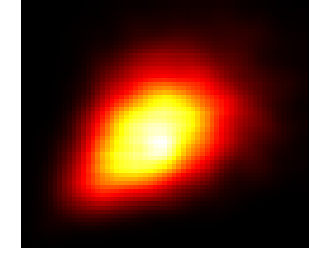}} &
  \raisebox{-.5\height}{\includegraphics{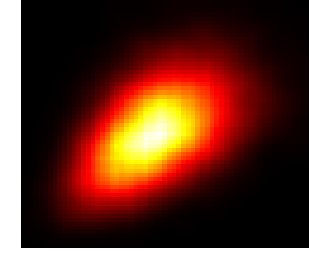}} \\

  D1 &
  \raisebox{-.5\height}{\includegraphics{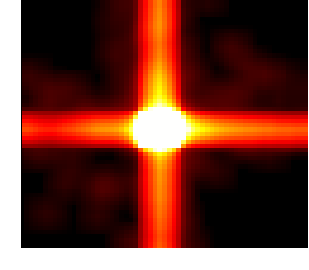}} &
  \raisebox{-.5\height}{\includegraphics{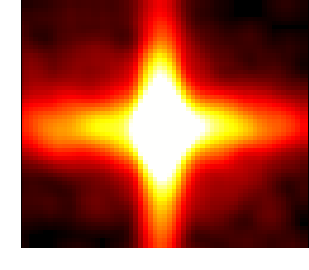}} &
  \raisebox{-.5\height}{\includegraphics{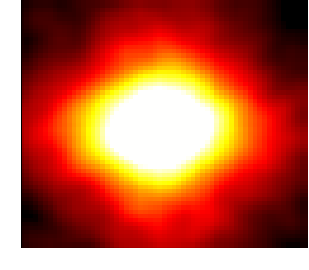}} &
  \raisebox{-.5\height}{\includegraphics{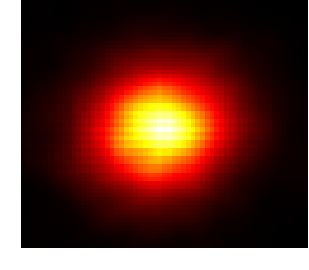}} &
  \raisebox{-.5\height}{\includegraphics{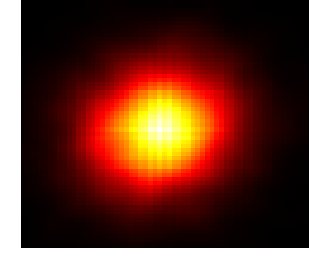}} \\

  D2 &
  \raisebox{-.5\height}{\includegraphics{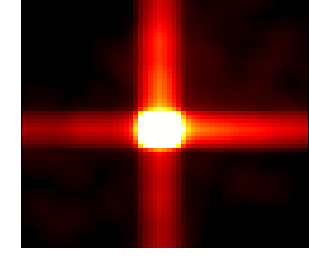}} &
  \raisebox{-.5\height}{\includegraphics{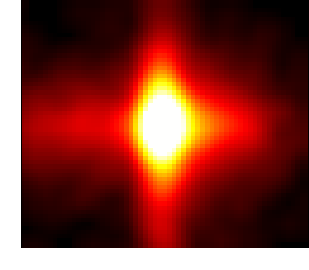}} &
  \raisebox{-.5\height}{\includegraphics{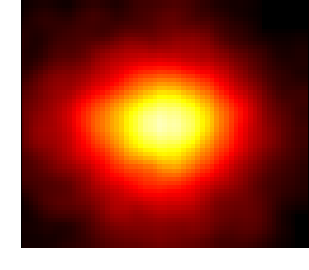}} &
  \raisebox{-.5\height}{\includegraphics{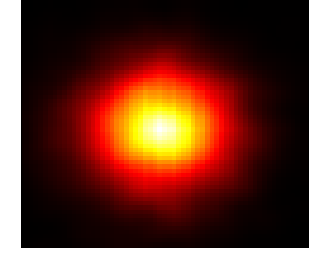}} &
  \raisebox{-.5\height}{\includegraphics{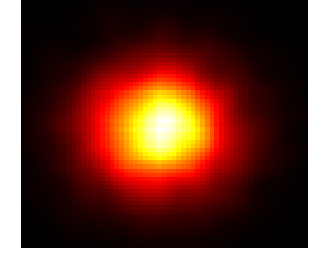}} \\

\end{tabular}

\caption{Smoothed histograms of pairwise module interactions. As the rewiring
  probability $p$ increases, interactions shift from nonlinear to linear. Time
  differencing acts as a highpass filter that adds stationarity to the data,
  but also affects the information content of the signal. Histograms are
  centered at the mean of each distribution and are $3\sigma$ wide.}

\label{fig:diffHist}

\end{figure*}

As we have argued above, the system exhibits a transition from competitive to
cooperative dynamics as coupling shifts from short- to long-range. This
transition is accompanied, in the large scale, by power law-like avalanche
statistics. In this section we explore the signatures of such transition in the
pairwise interactions between modules in different frequency bands.

To study the spectral aspects of the interaction, we analyse the data under
three filtering conditions:

\begin{description}
  \item[D0] Unfiltered data.
  \item[D1] After first-order differencing ($X'_t = X_t - X_{t-1}$).
  \item[D2] After second-order differencing ($X''_t = X'_t - X'_{t-1}$).
\end{description}

Since time-differencing is essentially a highpass filter, by taking successive
differences we are effectively exploring higher regions of the network's
frequency spectrum.

Figure \ref{fig:diffHist} shows aggregated histograms of the activity of all
pairs of modules for growing values of $p$ and the three filtering conditions.

For high $p$ the pairwise interactions visually appear Gaussian, suggesting
that interactions are mostly linear in this regime. For lower $p$, however, the
WTA dynamics are clearly visible and interaction is heavily nonlinear.
Interestingly, the transition between linear and nonlinear interaction lies in
the $p \in (0.2, 0.4)$ range, where information processing is most diverse and
avalanches exhibit power law-like statistics.

As a rough quantitative measure for nonlinearity, we calculate how much
information is accounted for by linear interactions. To do so we compare MI
between modules using two methods: the nonparametric KSG estimator, and a
parametric estimator under the assumption that all interactions are linear with
Gaussian noise.

The latter is referred to as the \emph{linear-Gaussian} estimator, and it
assumes that all variables in the system are jointly distributed as a
multivariate Gaussian distribution. In this case all relevant
information-theoretic quantities can be calculated analytically from the joint
covariance matrix of the system \cite[Chapter~9]{Cover2006}. Under this
assumption, the nonlinear component of the interaction is ignored.

To illustrate the effect of this assumption-breaking on informational measures,
in Fig.~\ref{fig:gausKSG} we show the average MI between all pairs of modules
(i.e. the MI between the two variables shown in the histograms of
Fig.~\ref{fig:diffHist}) calculated with the linear-Gaussian estimator and with
the nonparametric KSG estimator.

{ \setlength\belowcaptionskip{-3ex}
\begin{figure}[ht]
  \centering
  \includegraphics{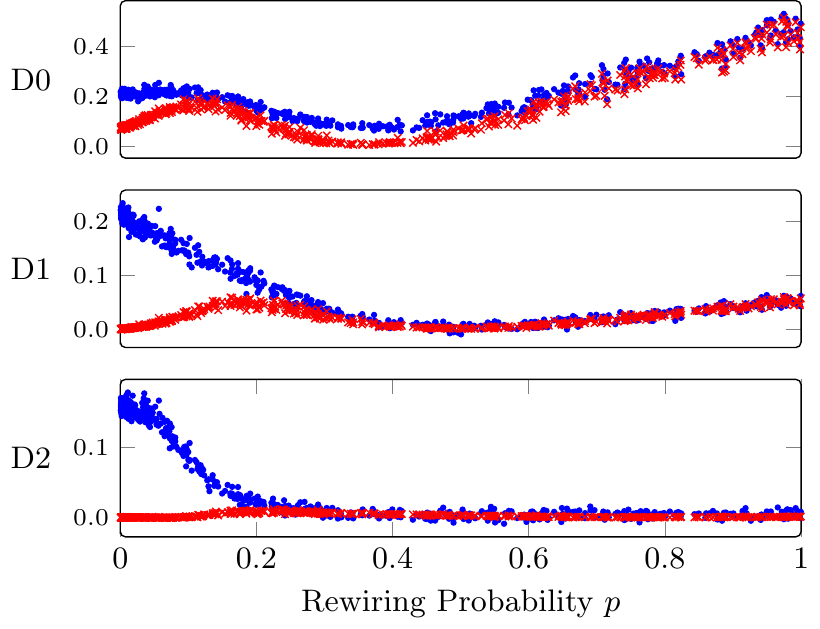}%
  \caption{MI between pairs of modules using linear-Gaussian (\textcolor{red}{red})
  and nonparametric (\textcolor{blue}{blue}) estimators.}%
  \label{fig:gausKSG}%
\end{figure}%
}%
As expected, the linear estimator always lies (up to random fluctuations) below
the KSG. By considering linear interactions only and ignoring the rest, linear
methods effectively provide a lower bound of the true MI. The linear-Gaussian
estimator is close to the KSG for high $p$ but consistently below it in the
low-$p$ range, which validates our claim that interactions shift from nonlinear
to linear with increasing $p$. Furthermore, the gap between both estimators is
more prominent in the differenced time series D1 and D2, indicating that the
linear component of the interaction is carried by lower frequencies, which are
more strongly suppressed by time-differencing.

\section{Conclusion}

In this paper we studied a simple modular spiking neural network and used it to
explore the relation between dynamics, information processing and underlying
network topology. The fully modular setting implements a WTA mechanism, whereas
the fully random setting is comparable to a PING oscillator -- both of which
are ubiquitous neural circuits in biological brains. This model gives us a way
of interpolating between the two in a continuous fashion by varying a long-range
connectivity parameter, $p$.

We find that for intermediate values of $p$ the network passes through a
near-critical regime in which avalanches display power law-like statistics,
with the same critical exponent as found in biological brains \cite{Beggs2003}.
Measures of information storage and transfer peak at either side of the
critical point, and the point where they are maximally balanced coincides with
the point where avalanches are closest to a power law.

This transition can also be understood as a breakdown of linearity, with
cooperative linear interaction being prevalent when connectivity is global and
delocalised, and competitive winner-take-all interaction more prominent when
connectivity is local.

Taken together, these findings link together three complementary views on
neural computation: topological, informational, and dynamical complexity.


\bibliography{ModularNetwork.bbl}

\end{document}